\documentclass[aps,prd,amsmath,amssymb,superscriptaddress,altaffillsymbol, preprintnumbers,preprint,nofootinbib,a4paper]{revtex4}
\pdfoutput=1
\usepackage{amsthm}
\usepackage{graphicx}
\usepackage{color}
\newcommand{\beq}{\begin{eqnarray}}
\newcommand{\eeq}{\end{eqnarray}}

\newcommand{\bmp}{\noindent\begin{minipage}{16cm}}
\newcommand{\emp}{\end{minipage}\vskip 7mm} 

\usepackage{dcolumn}
\usepackage{bm}
\usepackage{bbm}
\usepackage{subfigure}
\usepackage{pxfonts}
\usepackage{slashed}

\theoremstyle{definition}

\theoremstyle{plain}

\usepackage{epsfig}

\usepackage[ margin=5pt, font=normalsize,labelfont=bf,justification=raggedright]{caption}

\usepackage{hyperref}
\definecolor{rossoCP3}{cmyk}{0,.88,.77,.40}
\definecolor{verdeCP3}{rgb}{0.09765625, 0.57421875, 0.1015625}
\definecolor{bluCP3}{rgb}{0, 0.23, 0.67}
\hypersetup{colorlinks, bookmarksopen, bookmarksnumbered,citecolor=verdeCP3, linkcolor=bluCP3, pdfstartview=FitH, urlcolor=rossoCP3}
\def\lsim{\mathrel{\rlap{\lower4pt\hbox{\hskip1pt$\sim$}}
    \raise1pt\hbox{$<$}}}                
\def\gsim{\mathrel{\rlap{\lower4pt\hbox{\hskip1pt$\sim$}}
    \raise1pt\hbox{$>$}}}                

\newcommand{\be}{\begin{eqnarray}}
\newcommand{\ee}{\end{eqnarray}}

\begin{document}
\title{\Large  \color{rossoCP3} ~~\\ Perturbative extension of the standard model \\ with \\ a 125 GeV Higgs boson and magnetic dark matter}

\author{Karin Dissauer}
\email{dissauer@cp3-origins.net} 
\affiliation{{\color{rossoCP3} CP$^{3}$-Origins} \& Danish Institute for Advanced Study {\color{rossoCP3} DIAS}, University of Southern Denmark, Campusvej 55, DK-5230 Odense M, Denmark}
\affiliation{\mbox{Institute of Physics, Karl-Franzens University of Graz, Universit\"atsplatz 5, A-8010 Graz, Austria}}
\author{Mads T. Frandsen}
\email{frandsen@cp3-origins.net} 
\author{Tuomas Hapola}
\email{hapola@cp3-origins.net} 
\author{Francesco  Sannino}
\email{sannino@cp3-origins.net} 
\affiliation{{\color{rossoCP3} CP$^{3}$-Origins} \& Danish Institute for Advanced Study {\color{rossoCP3} DIAS}, University of Southern Denmark, Campusvej 55, DK-5230 Odense M, Denmark}

\begin{abstract}
We introduce a perturbative extension of the standard model featuring a new dark matter sector together with a 125 GeV Higgs.  The new sector consists of a vectorlike heavy electron $(E)$, a complex scalar electron $(S)$ and a standard model singlet Dirac fermion $(\chi)$.  The interactions among the dark matter candidate $\chi$ and the standard model particles occur via loop-induced processes involving the operator $S\overline{E} \chi$y, with $y$ being the Yukawa-like coupling. 
The model is an explicit underlying realization of  the light magnetic dark matter effective model introduced earlier to alleviate the tension among several direct dark matter search experiments.  We further constrain the parameters of the underlying theory using results from the Large Hadron Collider. The extension can accommodate the recently observed properties of the Higgs-like state and leads to interesting predictions.  Finally we show that the model's collider phenomenology and constraints nicely complement  the ones coming from dark matter searches. 
\\[.1cm]
{\footnotesize  \it Preprint: CP$^3$-Origins-2012-31, DIAS-2012-32.}
 \end{abstract}

\maketitle

\section{A {$S\overline{E}\chi \lowercase{y}$} extension of the standard Model}
After the recent discovery \cite{:2012gk,:2012gu} of a new resonance with properties similar to the standard model (SM) Higgs particle, it is particularly timely to investigate minimal extensions of the SM featuring also dark matter sectors. Here we put forward a renormalizable extension featuring, besides the Higgs at 125 GeV, also a dark matter sector.   The new sector consists of a vectorlike heavy electron ($E$), a complex scalar electron ($S$) and a SM singlet Dirac fermion ($\chi$). The associated renormalizable Lagrangian is 
\begin{eqnarray}
\label{Sexy}
    \mathcal{L}_{S\overline{E}\chi {\rm y}}& =& \mathcal{L}_{\text{SM}}+\bar{\chi}i\slashed{\partial}\chi - m_\chi\bar{\chi}\chi + \overline{E}i\slashed{D}E - m_E\overline{E}E - (S \overline{E}{\chi}  y + \text{H.c.} ) \nonumber \\
& +& D_\mu S^{\dagger}D^\mu S - m_S^2S^{\dagger}S-\lambda_{HS}H^{\dagger}HS^{\dagger}S - \lambda_S (SS^{\dagger})^2\ ,\end{eqnarray}
where $D^\mu=\partial^\mu-ie\frac{s_w}{c_w}Z^{\mu}+ieA^\mu$, $s_w$ and $c_w$ represent the sine and cosine of the Weinberg angle. 
We assume the new couplings $y, \lambda_{HS}$ and $ \lambda_S$ to be real and the bare mass squared of the $S$ field to be positive  so that the electroweak symmetry breaks via the SM Higgs doublet ($H$). The interactions among $\chi$, our potential dark matter candidate, and the SM fields occur via loop-induced processes involving the $S\bar{E} \chi$y operator in \eqref{Sexy}.
If, however, the Higgs sector of the SM is natural and described, for example, by  composite dynamics \cite{Sannino:2009za}, one can imagine the new sector, and, in particular, the scalar $S$, also to be composite.

At the one-loop level, $\chi$  develops the following magnetic-type interactions\begin{equation}
\mathcal{L}_5 = \frac{\lambda_\chi}{2}\bar{\chi}\sigma_{\mu\nu}\chi F^{\mu\nu}- \frac{s_w}{2 \,c_w}\lambda_\chi\bar{\chi}\sigma_{\mu\nu}\chi Z^{\mu\nu},
\label{eq:mmi}
\end{equation}
where $F^{\mu\nu}$ and $Z^{\mu\nu}$ are the photon and $Z$ field strength tensors, and $\lambda_\chi$ is  related to the electromagnetic form factor $F_2(q^2)$, 
\begin{equation}
\lambda_\chi = \frac{F_2(q^2)}{2m_\chi}e \ .
\end{equation}
 The explicit derivation of the one-loop--induced form factor can be found in Appendix \ref{ap:1}.  It is instructive to report the analytic form for a few interesting limits to learn about the dependence upon couplings and masses. We start by considering the static limit $F_2(0)$, useful for dark matter direct detection experiments.
To reduce the parameter space we take the masses of $E$ and $S$ to be degenerate, $m_E=m_S=M$, leading to the simplified expression
\begin{equation}
F_2(0) = \frac{y^2}{8\pi^2}\left(  \frac{2}{z}\sqrt{\frac{2+z}{2-z}}\tan^{-1}\left(\frac{z}{\sqrt{4-z^2}}\right) -1 \right),
\end{equation} 
where $z=\frac{m_{\chi}}{M}$. For small $z$ we have
\begin{equation}
F_2(0)= \frac{y^2}{16\pi^2}\left( z + \frac{z^2}{3} \right) + \mathcal{O}\left( z^3 \right) \ ,
\end{equation} 
which for $z=0$ gives $\lambda_\chi = {ey^2}/(32\pi^2 M)$.  It is also interesting to consider a dark matter candidate with mass  of the order of the electroweak scale or slightly higher. For this purpose a simple estimate for the electromagnetic form factor can be deduced by setting $z = 1$ (i.e. $m_{\chi} =M$), and $M$ around the electroweak scale, yielding $\displaystyle{F_2(0)=y^2\frac{\sqrt{3}\pi-3}{24\pi^2}}$ and, therefore, $\displaystyle{\lambda_\chi =e y^2\frac{\sqrt{3}\pi-3}{48\pi^2 M}} $. From the two limits it follows that the scale of the magnetic moment is controlled by the common mass of the heavy states $M$.

The model allows us to investigate the interplay between dark matter and ordinary matter and  the possibility that dark matter is either light with respect to the electroweak scale or has the same mass scale.  The model can be constrained by dark matter experiments and at the (LHC).  

The light dark matter limit allows us to explore models of long-range interactions which can be useful to alleviate the tension between the experimental observations by DAMA/LIBRA \cite{Bernabei:2008yi} and the limits set by XENON100 \cite{Aprile:2011hi,Aprile:2012nq} and CDMS \cite{Ahmed:2009zw}. The light dark matter limit naturally maps in the effective model  of magnetic light dark matter put forward in, e.g., Refs. \cite{An:2010kc,Banks:2010eh,DelNobile:2012tx,Fitzpatrick:2010br}, while allowing us to investigate its phenomenology at the electroweak energy scale. 

{Electroweak scale dark matter is potentially observable in cosmic-ray detection experiments.} If the {tentative} observation of the 135 GeV Fermi line \cite{Bringmann:2012vr,Weniger:2012tx,Huang:2012yf,Hektor:2012ev} is {confirmed} by future data, the model studied here can be opportunely extended to explain it \cite{Weiner:2012gm}. The present model, albeit similar to the one of Ref. \cite{Weiner:2012gm}, was conceived independently in order to provide an UV completion of the analysis presented in Ref. \cite{DelNobile:2012tx}. In addition the main differences with the model presented in Ref. \cite{Weiner:2012gm} are: Our model describes elastic magnetic dark matter; we discuss all the relevant renormalizable interactions with the SM, in particular, the coupling of the new sector to the SM fermions  and the Higgs. The coupling to the Higgs is quite important, allowing us to properly investigate the collider phenomenology of the model.

The paper is organized as follows. In Sec. \ref{lmix} we allow for the heavy electron to decay directly into SM fermions by adding mixing operators with the SM leptons. We will set preliminary bounds on the mixing angles.  In Sec. \ref{direct} we investigate collider constraints on the spectrum and couplings of the $S\overline{E}\chi$y sector. In particular, we will focus on the invisible width of the $Z$ and the Higgs alongside the decay of the latter into $\gamma\gamma$ and $\gamma Z$. If dark matter is sufficiently light to be produced at the LHC, it can affect missing energy signals \cite{Chatrchyan:2012me} which will be also explored for this model. In Sec. \ref{mdmp} we summarize the light dark matter constraints set in Ref. \cite{DelNobile:2012tx} and translate them in terms of the fundamental parameters of the model.

\section{Allowing $E$ to decay into SM leptons}
\label{lmix}

{Heavy single charged leptons cannot be stable on cosmological scales because they would appear as anomalously heavy isotopes. The abundance of such isotopes is limited to be less than one in $10^{12}$ baryons \cite{Nitz:1986gb} as discussed in, e.g., Refs.  \cite{Cahn:1980ss,Chivukula:1989qb}.}
 Therefore one needs to mix the heavy electron with the SM leptons. Here we assume, for simplicity, that the mixing is primarily with the $\tau$ lepton.  In general, we can also mix $\chi$ with the SM neutrinos provided that its decay time is sufficiently long on cosmological scales to be a proper dark matter candidate. As a first approximation, we take $\chi$ to be stable and, henceforth, suppress its mixing with SM neutrinos. 


The mixing Lagrangian, after the Higgs field has acquired a vacuum expectation value $v \simeq 246$~GeV, is
\begin{equation}
{\cal L}_{E\tau} = \bar{L}'~M~R'=
\begin{pmatrix}
\bar{\tau}'_L & \bar{E}'_L
\end{pmatrix}
\begin{pmatrix}
y_\tau\frac{v}{\sqrt{2}} &  y_E\frac{v}{\sqrt{2}} \\
m & m_E
\end{pmatrix}
\begin{pmatrix}
\tau'_R \\ E'_R
\end{pmatrix} .
\end{equation}
 Here the prime on the fields indicates gauge eigenstates. The existing SM mass term for the $\tau$ lepton is the first entry of the matrix. The remaining off-diagonal terms are added to \eqref{Sexy}.
 
To achieve the mass eigenstates, we need to diagonalize $M$ which is a nonsymmetric real matrix. The diagonalization procedure makes use of two independent rotations, one for the left-handed fields $V$ and the other for the right-handed fields $U$. The diagonalized matrix is then
\begin{equation}
M_D = V^{\top}~M~U\ ,
\end{equation}
where $V$ and $U$ are the following real orthogonal matrices 
\begin{equation}
V=
\begin{pmatrix}
\cos\theta & \sin\theta \\
-\sin\theta & \cos\theta
\end{pmatrix}\ , \qquad 
U=
\begin{pmatrix}
\cos\alpha & \sin\alpha \\
-\sin\alpha & \cos\alpha
\end{pmatrix} \ .
\end{equation}
$V$ and $U$ 
diagonalize, respectively, the symmetric squared mass matrices $M M^\top$ and $M^\top M$.  The mixing angles are given by the relations
\begin{align}
\tan 2\alpha = - \frac{\sqrt{2} v (m y_\tau + m_E y_E)}{\frac{v^2}{2} (y_\tau^2+ y_E^2)-m^2-m_E^2} \ ,  \\
\tan 2\theta = - \frac{2(y_\tau y_E \frac{v^2}{2} +m m_E)}{m^2-m_E^2+\frac{v^2}{2}(y_\tau^2-y_E^2)} \ .
\end{align}
The physical mass of the $\tau$ lepton fixes one of the eigenvalues. Constraints on the left-handed mixing angle  $\theta$ come from the $Z$ decay and from the electroweak precision measurements as discussed in Ref. \cite{Popovic:2000dx}. These constraints lead to an upper limit, which at the 95\% confidence level is
\begin{equation}
\sin^2\theta < 0.0018 \ .
\end{equation} 
Depending on the values of the (yet unknown) physical parameters, the right-handed angle $\alpha$ can be larger or smaller than $\theta$. In the future, an interesting way to further constrain the right-handed angle would be by measuring the decay of the Higgs to $\tau\tau$. 

To simplify the analysis here, we demand $E$  to be stable on collider detection time scales. This requires the mixing angles to be of the order of
\begin{equation}
\sin^2\theta \sim \sin^2\alpha \sim 10^{-14},
\end{equation}
rendering the mixing effects negligible\footnote{However, if the mixing is large, it will lead to interesting phenomenological consequences which we plan to investigate elsewhere while we mention here a few.  For example, the $S\overline{E}\chi y$ coupling leads to the following mixing operators between a generic SM lepton,  $\chi$, and the scalar, $S$ :
\begin{equation}
S(\bar{E}_L\cos\theta+\bar{\ell}_L\sin\theta)\chi y + S(\bar{E}_R\cos\alpha+\bar{\ell}_R\sin\alpha)\chi y + \text{H.c.},
\end{equation}
inducing a correction to the $g-2$ of the given SM lepton $\ell$ as well as a modification of its electric dipole moment. }.  

However, in this limit at least one of the new charged particles is stable and will hit, once produced, the walls of the LHC detectors. The CMS Collaboration excluded masses below $427$ GeV for particles with one $e$  charge \cite{CMS:HSCP}. The limit is imposed for a lepton with SM-like neutral-current interactions. If we interpret the CMS results in the context of our model, we find that $E$ and $S$ have to be more massive than $393$ GeV and $243$ GeV, respectively.

\section{Constraints from the Large Hadron Collider}
\label{direct}

Having at our disposal the explicit Lagrangian for this particular extension of the SM, we can now systematically analyze the various phenomenological constraints and potential signals at the LHC. 

\subsection{Invisible width of the Higgs}
\begin{figure}[htp]
  \centering
  \includegraphics[width=0.7\textwidth]{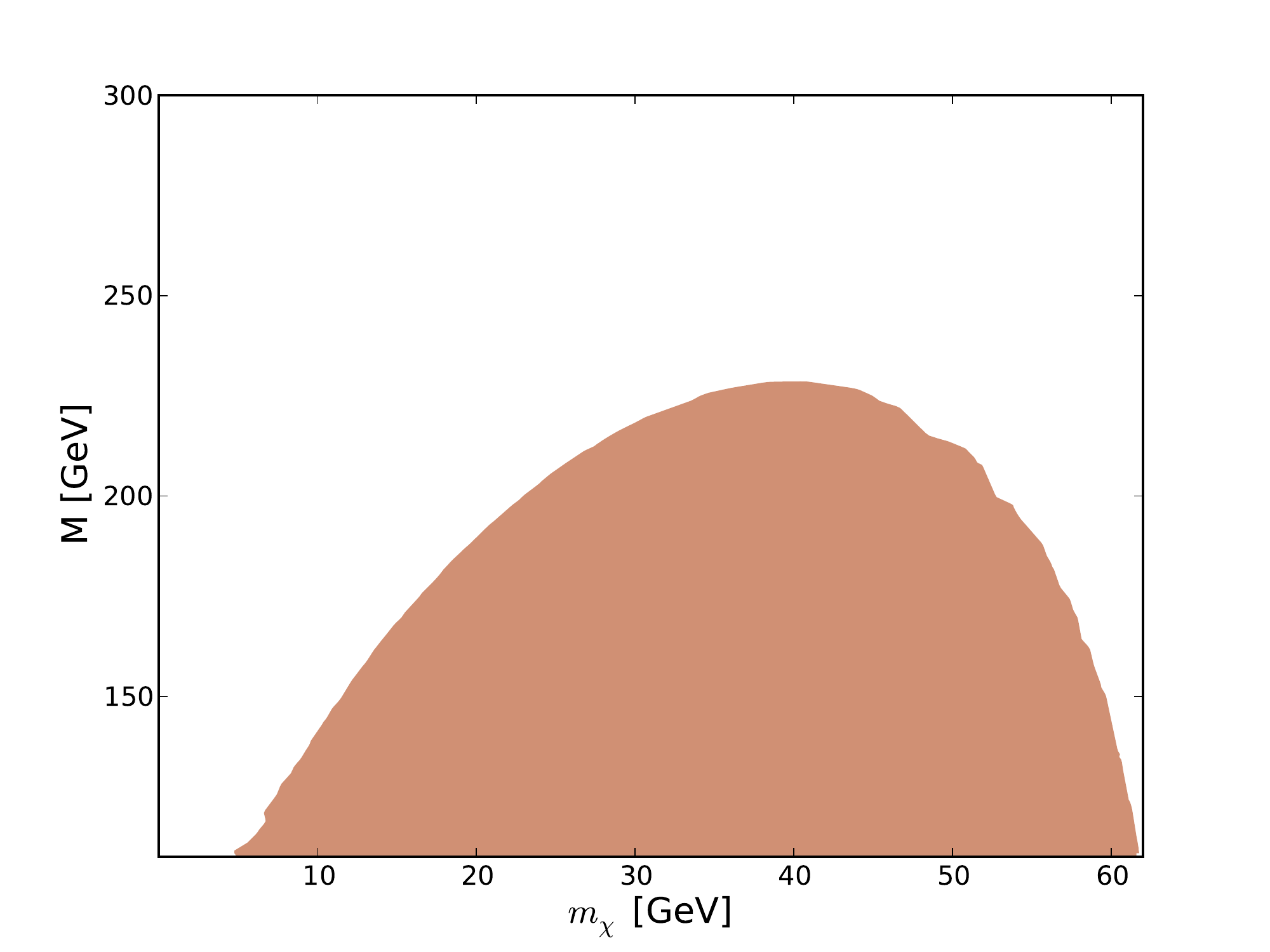}
  \caption{The  shaded region of the parameter space  ($m_\chi$, $M$)  assuming $\lambda_{HS}=y=4\pi$ is excluded at the 1$\sigma$ level.  }
  \label{fig:hinvis}
\end{figure}

Via loop processes the Higgs can decay into two $\chi$ fermions giving a contribution on the invisible width of the Higgs. The explicit computation and relevant formulas are derived in Appendix \ref{InvisibleH}. Of course, for this process to occur, $\chi$ has to have a mass less than or around half of the mass of the Higgs. We assume that the dominant contribution comes from the decay of the Higgs into two $\chi$s, and, therefore, the invisible width  is
\begin{equation}
\Gamma_{\rm inv}[H] \approx \Gamma[H\to 2\chi] =\frac{m_H}{32\pi}\left( 4 v^2\lambda_{HS}^2y^4 |A|^2  \right)\left( 1-\frac{4m_{\chi}^2}{m_H^2} \right)^{\frac{3}{2}}.
\end{equation}
The loop function $A$ is given in Appendix \ref{InvisibleH}. From the results of Ref. \cite{Giardino:2012dp}, one deduces the rough $1\sigma$ estimate
\begin{equation}
\frac{\Gamma[H\to 2\chi]}{\Gamma_{\text{SM}}+ \Gamma[H\to 2\chi] } \leq 0.15 \ . 
\end{equation}
In Fig. \ref{fig:hinvis} we present the associated disfavored parameter space in the ($m_\chi$, $M$) plane assuming very large values of $\lambda_{HS}=y=4 \pi$ to maximize the constraints\footnote{These large values of the couplings stretch the validity of perturbation theory and are taken for illustrative reasons. It would be interesting to extend the perturbative analysis to higher orders. }.
 
\subsection{Invisible width of the Z}
 If $\chi$ has a mass below or around half of the $Z$ mass, then one can use the invisible width of the $Z$ as a further constraint. The measured invisible width of the $Z$ is \cite{Beringer:1900zz}
 \begin{equation}
 \Gamma^Z_{\text{inv}}=499.0\pm1.5~\text{MeV} \ ,
 \end{equation}
 yielding a 95\% confidence level upper limit \cite{Carena:2003aj}
 \begin{equation}
 \label{eq:invz}
 \Gamma_{\text{inv}}^{\text{new}} < 2.0 ~\text{MeV}
 \end{equation}
  on the invisible width contribution from new physics which does not interfere with the neutrino pair production.
  
\begin{figure}[htp]
  \centering
  \includegraphics[width=0.7\textwidth]{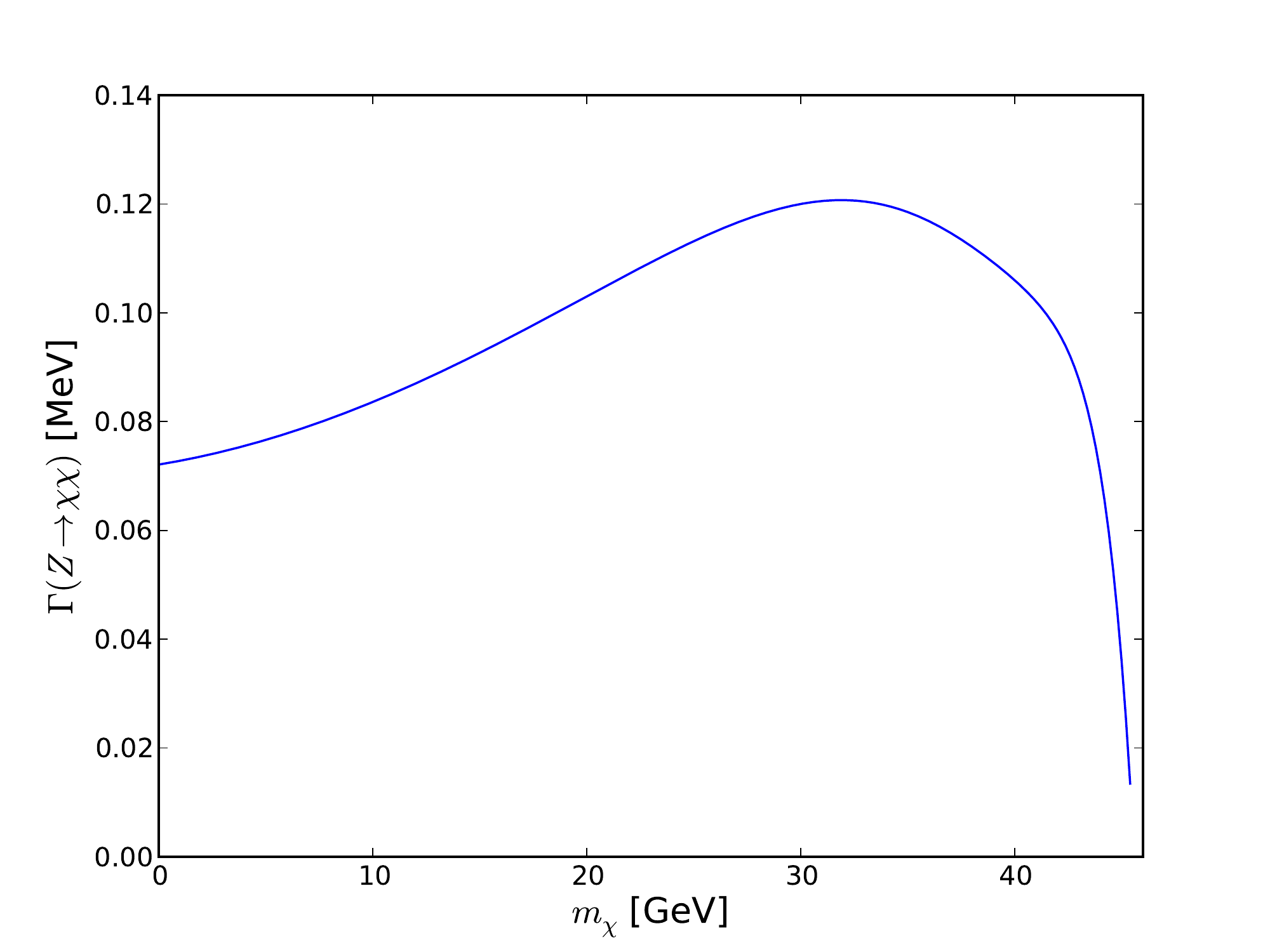}
  \caption{$Z$ decay width to a $\chi$ pair as a function of $m_\chi$ taking $y=4\pi$ for the $S\overline{E}\chi y$ coupling.}
  \label{fig:zinvis}
\end{figure}

 The $Z\to\chi\bar{\chi}$ decay width formula is
  \begin{equation}
  \begin{split}
  \Gamma^{Z\to\chi\bar{\chi}}=&\frac{e^2s_w^2}{96 c_w^2 m_\chi^2 m_Z\pi}\sqrt{1-\frac{4m_\chi^2}{m_Z^2}} \\
  &\left( 24m_\chi^2m_Z^2F_1(m_Z^2) F_2(m_Z^2)+8F_1^2(m_Z^2)m_\chi^2\left(2m_\chi^2+m_Z^2\right)+F_2^2(m_Z^2)m_Z^2(8m_\chi^2+m_Z^2) \right).
  \end{split}
  \end{equation}
It is clear from Fig. \ref{fig:zinvis} that the invisible constraint on the $Z$ boson decay width into $\bar{\chi}\chi$  is weak even for large values of the coupling $y=4\pi$. 
  
 \subsection{Monojets}

\begin{figure}[h!]
  \centering
  \includegraphics[width=0.4\textwidth]{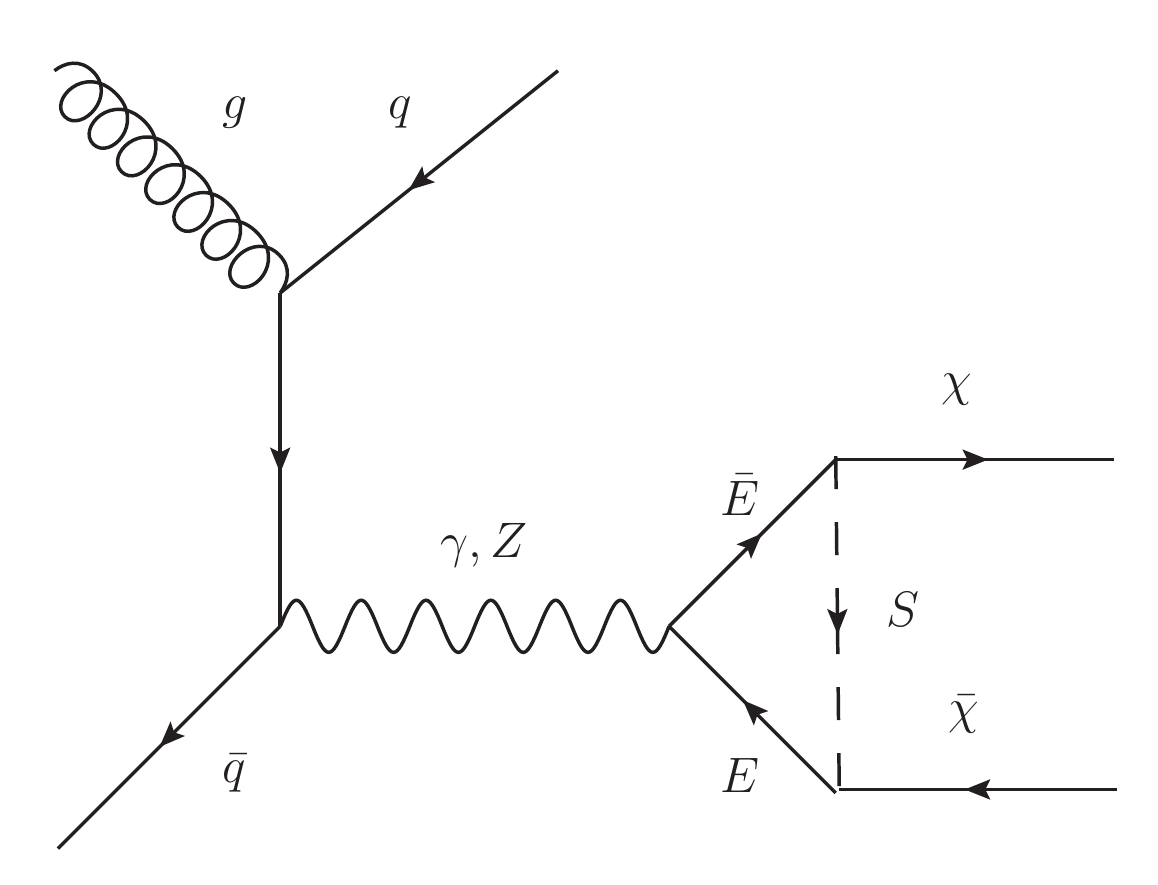}
    \includegraphics[width=0.4\textwidth]{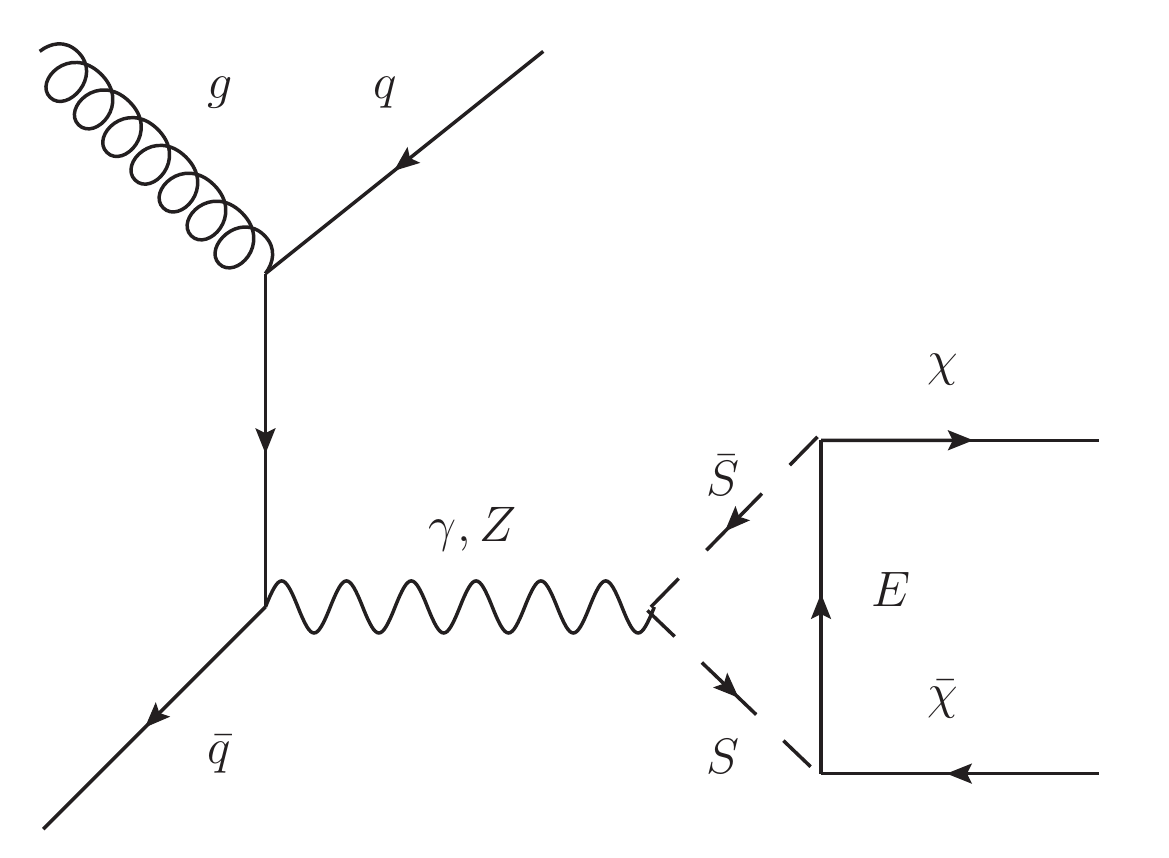}
  \caption{Two diagrams giving a monojet signature at one-loop level through \mbox{quark-gluon} initial states. Similar diagrams arise also via \mbox{quark-quark} initial states.}
  \label{fig:monojet}
\end{figure}

\begin{figure}[ht!]
  \centering
  \includegraphics[width=0.7\textwidth]{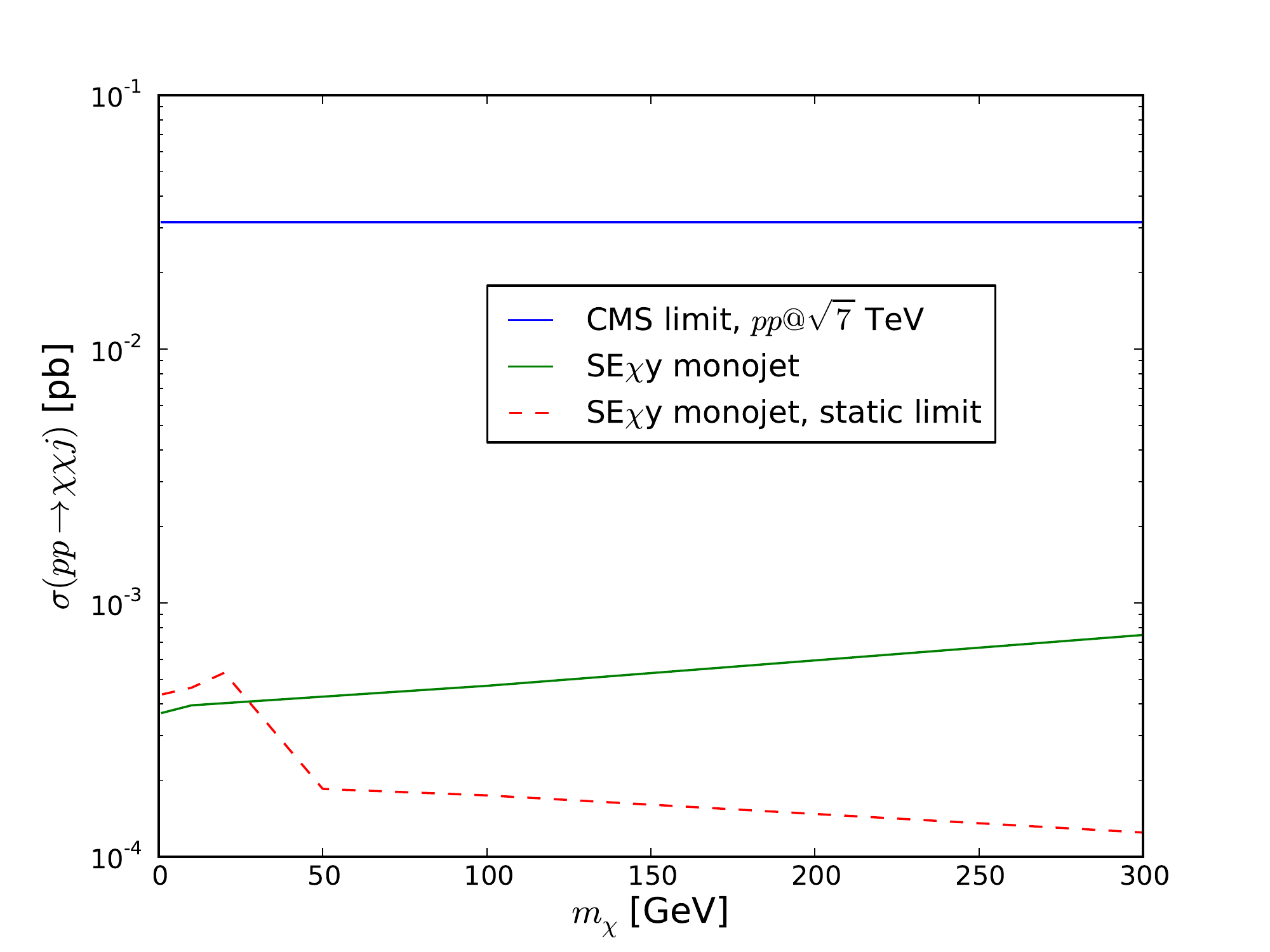}
  \caption{The lower line: monojet cross section as a function of $m_\chi$ using parameter values $M=300$ GeV and $y=4\pi$. The upper line: CMS collaboration $95\%$ confidence level upper limit on the monojet signal cross section.}
  \label{fig:F2}
\end{figure}

 Monojet signatures, associated to missing energy, can emerge in this model from diagrams of the kind reported in Fig. \ref{fig:monojet}.
Analytic expressions for the corresponding differential cross sections in the center-of-mass frame are given in Appendix \ref{monoappendix}. We have implemented the momentum-dependent loop functions given in Appendix \ref{ap:1} into \texttt{MadGraph5} \cite{Alwall:2011uj}. This allows us to compute the monojet cross section and compare it with the experimental results. We take here the common mass for the new charged particles to be $M=300$ GeV and the coupling $y$ to be $4\pi$. In Fig. \ref{fig:F2} we have plotted the cross section for a monojet signal as a function of $m_\chi$ and the $95\%$ confidence level upper limit set by the CMS Collaboration \cite{Chatrchyan:2012me}.
From the plot it is clear that the coupling $y$ needs to be at least 3 times larger than $4\pi$ to reach the current experimental sensitivity. Thus, the monojet searches do not constrain yet the  model parameters.

Monojets arising through a magnetic moment interaction have been studied in Refs. \cite{Barger:2012pf,Fortin:2011hv}. In these papers the magnetic moment $\lambda_\chi$  is taken to be a momentum-independent constant, and limits on its value are derived. In the static limit and for $m_\chi \approx 10$ GeV, we recover their results when taking $y \approx 4\pi$ and $M \approx 10$ GeV.

\subsection{Direct LHC probe of the $S$ sector}
The scalar $S$ has properties reminiscent of a selectron except that the heavy electron is vectorlike, and, therefore, only the scalar $S$ feels the Higgs directly. This is true provided we do not mix the new electron with the SM leptons via generalized Yukawa interactions. Due to this property and the requirement of the renormalizability of the theory, the $S$ sector can be probed directly using processes involving the Higgs. Here we will consider the Higgs to two neutral gauge bosons processes.

\subsubsection{$h\to \gamma\gamma$} 

The measured signal strength in the Higgs to a two-photon decay channel is slightly bigger than the expectation from the SM \cite{ATLAS:di,CMS:di}. A nice feature of this model is that it can explain the enhancement via the  $S$ contribution to this process. The enhancement fixes the sign of the coupling $\lambda_{HS}$ once the SM Higgs to top coupling is employed.

The production of the Higgs is not altered compared to the SM. This allows us to write the signal strength relative to the SM one as
\begin{equation}
\mu=\frac{\sigma(pp\to h\to\gamma\gamma)}{\sigma_{{\rm SM}}(pp\to h\to\gamma\gamma)}=\frac{\Gamma_{{\rm SM}}}{\Gamma_{{\rm SM}}-\Gamma_{{\rm SM}}^{h\to\gamma\gamma}+\Gamma_{S\overline{E}\chi y}^{h\to\gamma\gamma}+\Gamma_{S\overline{E}\chi y}^{h\to\chi\chi}}\frac{\Gamma_{S\overline{E}\chi y}^{h\to\gamma\gamma}}{\Gamma_{{\rm SM}}^{h\to\gamma\gamma}},
\end{equation}
where $\Gamma_{X}^{h\to\gamma\gamma}$ denotes a partial width and $\Gamma_{{\rm SM}}$ the total width. The partial width of the Higgs into two $\chi$s depends on the physical value of the $\chi$ mass, which here we take to be about $10$~GeV. 
The needed formulas to calculate the Higgs decay width to two photons are given in Appendix \ref{hgamma}. The total width of the Higgs is calculated with the aid of \texttt{HDECAY}  \cite{Djouadi:1997yw}. 

\begin{figure}[ht!]
  \centering
  \includegraphics[width=0.49\textwidth]{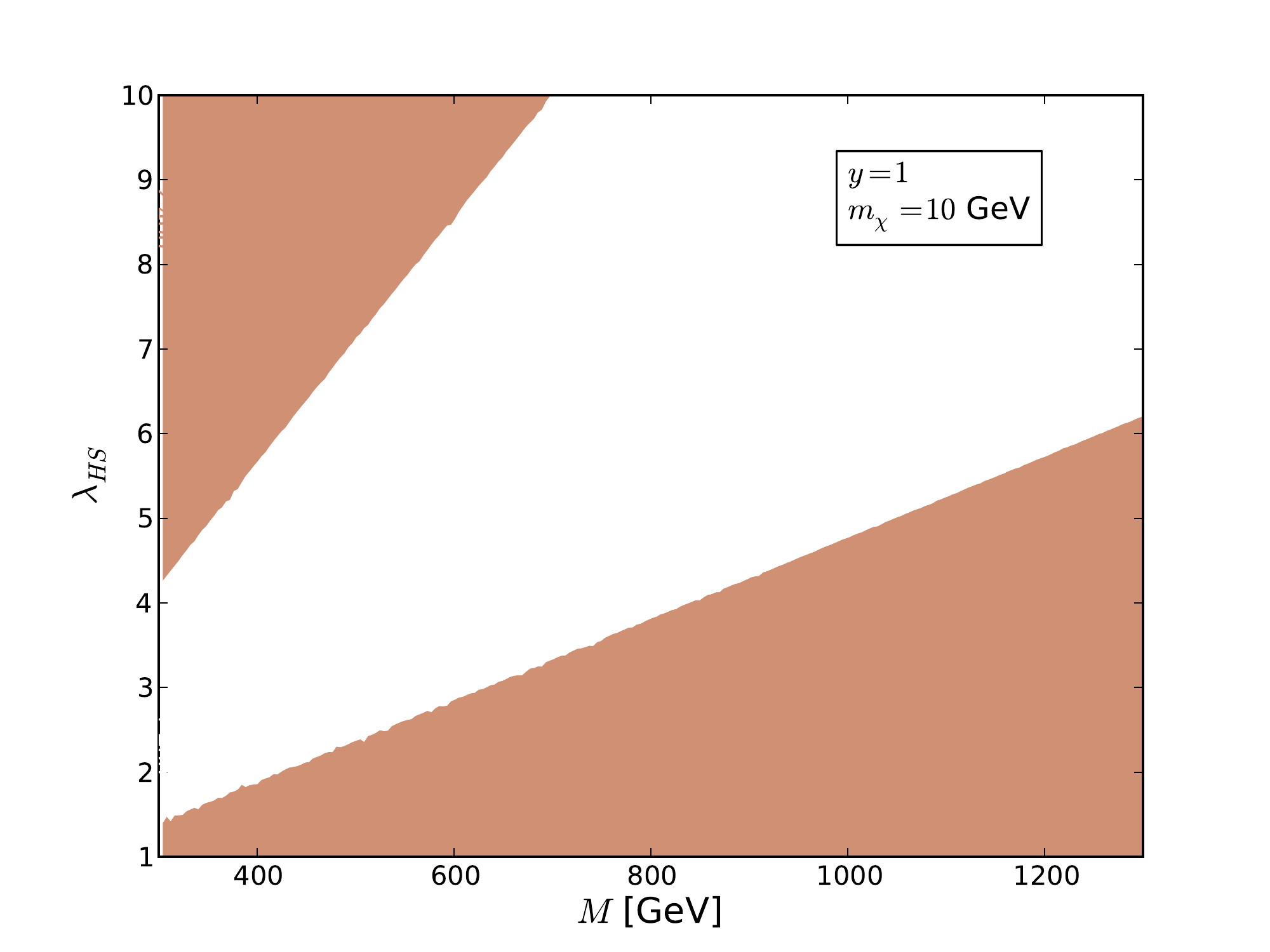}  \includegraphics[width=0.49\textwidth]{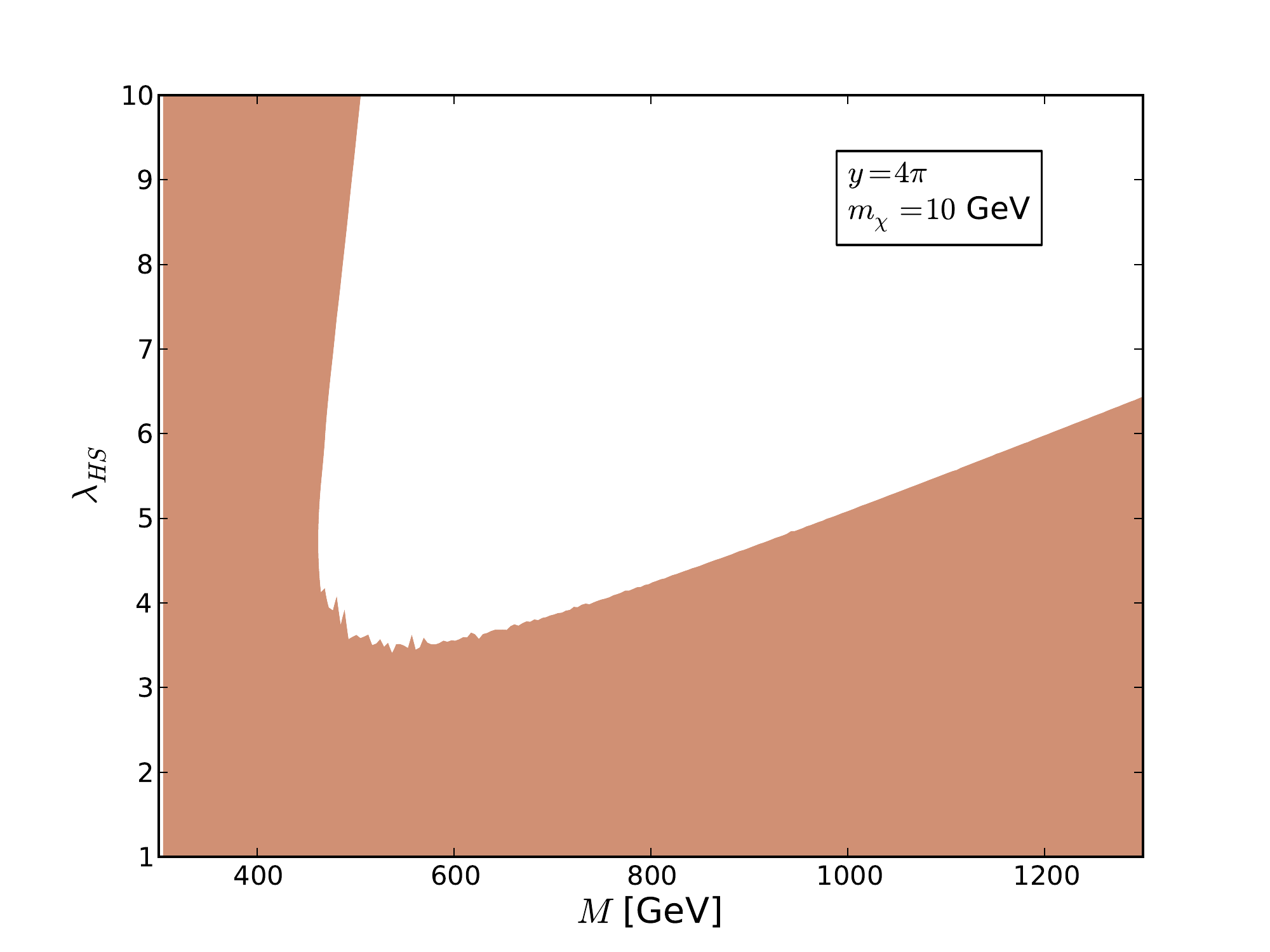}
  \caption{The shaded areas represent the excluded parameter space  ($M$, $\lambda_{HS}$) with $m_{\chi} = 10$~GeV based on the ATLAS collaboration measurement of the $h\to \gamma\gamma$ signal strength.}
  \label{fig:hdec}
\end{figure}

The best-fit value for the signal strength reported by the ATLAS Collaboration \cite{ATLAS:di} is
\begin{equation}
\mu_{\text{ATLAS}}=1.9\pm0.5,
\end{equation}
achieved by combining the $\sqrt{s}=7$ TeV and $\sqrt{s}=8$ TeV data samples of size 4.8 fb$^{-1}$ and 5.9 fb$^{-1}$, respectively. In the context of our model, we can test which pair of values of $(M,\lambda_{HS} )$ gives $\mu_{\rm ATLAS}$ within the quoted error. In Fig. \ref{fig:hdec} the shaded areas represent excluded parameter combinations based on this ATLAS measurement. 

\subsubsection{$h \to \gamma Z$}
 The model can be tested further when the process $h \to \gamma Z$ will be measured. A thorough analysis of how a singlet scalar, or other non-SM particles, affects the $h\to\gamma Z$ decay is performed in Ref. \cite{Carena:2012xa}. Adapting their analysis to our model shows that if the diphoton decay is enhanced, then the $\gamma Z$ decay will also be enhanced. This is so since $S$ is a weak isosinglet. The situation changes when $S$ is charged under weak  isospin interactions since  the $Z$ couplings are sensitive  to the full $SU(2)_{L}\times U(1)$ quantum numbers. From an experimental viewpoint this process is more challenging than the analogous two-photon decay process.  This is so since the $Z$ boson further decays and the large QCD background allows only the leptonic decay modes to be studied, which, however, have small branching ratios of the order of 3.3\%.

\section{Summarizing Collider and Dark Matter constraints }
\label{mdmp}
Our dark matter candidate is a Dirac fermion $\chi$ interacting via a magnetic dipole moment, given in \eqref{eq:mmi}. Depending on the value of its mass and the size of the interactions, it can appear in different dark matter search experiments.

For the light dark matter case, i.e.~the mass of $\chi$ in the tens of GeV region which is also interesting for collider physics, it has been shown that the model leads to interesting phenomenological consequences \cite{DelNobile:2012tx} which we briefly summarize here. This is due prevalently to the long-range nature of the dark matter interactions with ordinary matter.

We concentrate on results from DAMA, CoGeNT, CRESST, as well as CDMS, XENON and PICASSO. It was shown in Ref. \cite{DelNobile:2012tx} that there is a region of parameter space able to alleviate the tension between the experiments when using  magnetic moment interactions. It was observed that one can, in fact, bring DAMA, CoGeNT and CRESST signals to overlap while being marginally consistent with CDMS, XENON and PICASSO experiments (see Fig.~3 of Ref.~\cite{DelNobile:2012tx}). 
A  best-fit result  leads to a dark matter mass around $10$~GeV and  a constant magnetic moment of about \mbox{$1.5 \times 10^{-18}$ $e$ cm}, which corresponds to  \mbox{$ 32 \pi^2 M/y^2 =\frac{e}{\lambda_\chi}\sim 10$ TeV}. For example, for $M\approx 500$~GeV we find $y \approx \pi$ meaning that the underlying dynamics of the model can be explored at the electroweak scale.  

\begin{figure}[h!]
  \centering
  \includegraphics[width=0.49\textwidth]{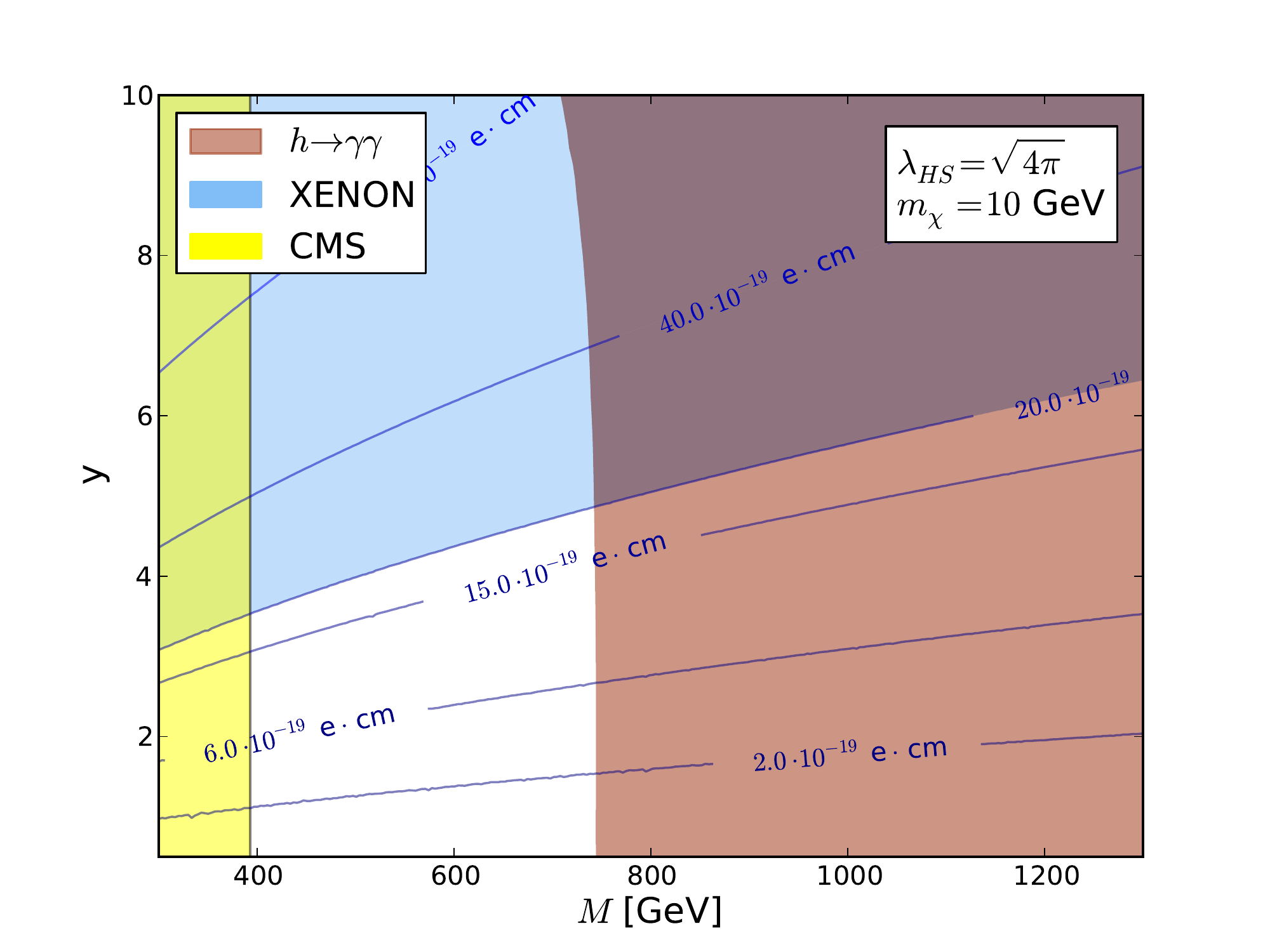}  \includegraphics[width=0.49\textwidth]{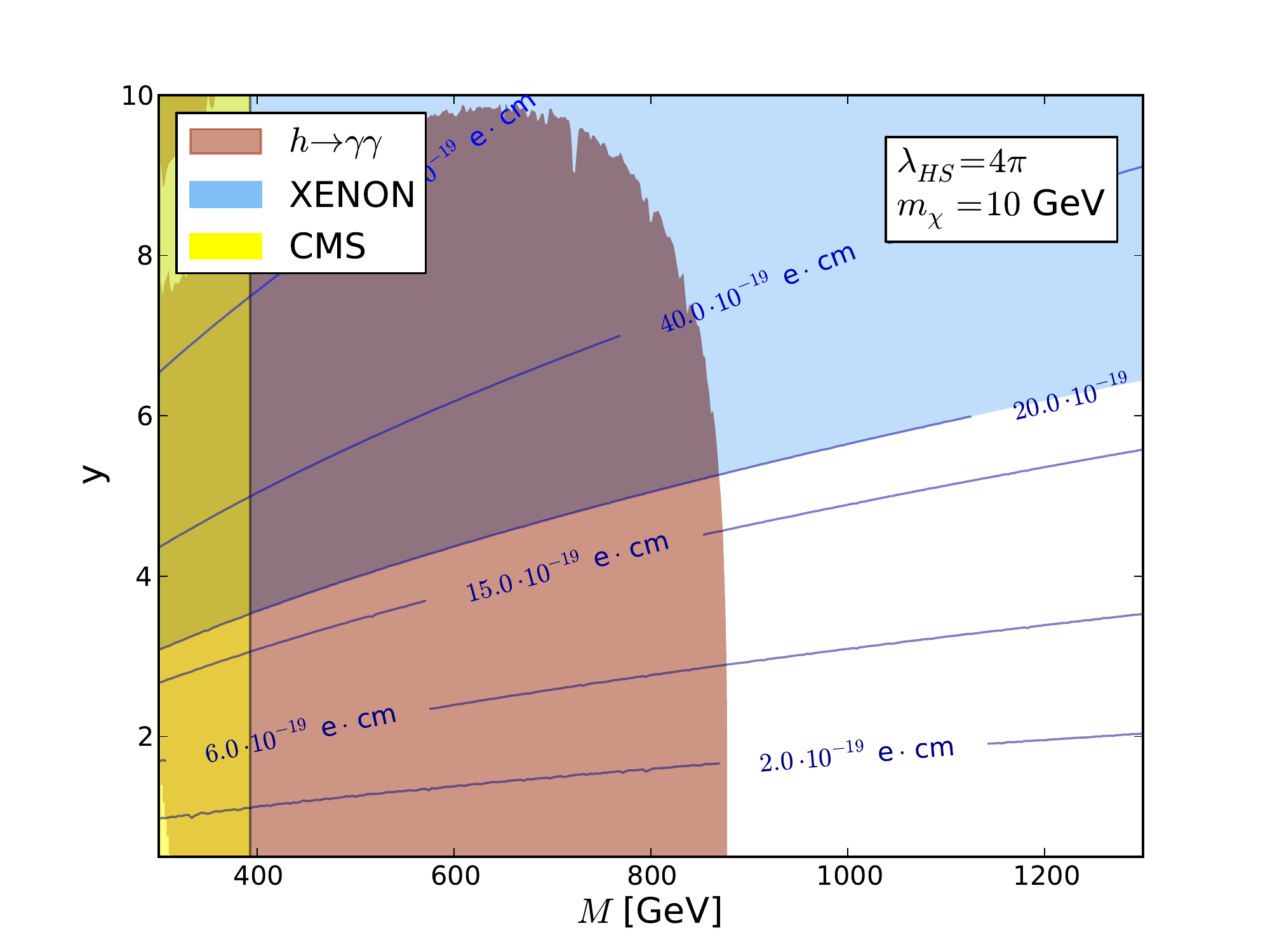}
  \caption{Strength of the magnetic moment $\lambda_\chi$ in the $(M,y)$ plane with $m_\chi = 10 $ GeV and for $\lambda_{HS}=\sqrt{4\pi}$  (left panel) and $\lambda_{HS}={4\pi}$  (right panel). The unshaded region is the one allowed by experiments. }
  \label{fig:mmoment}
\end{figure}

The contour plot of equal $\lambda_\chi$ in the ($M$,$y$) plane is shown in Fig.~\ref{fig:mmoment} together with the exclusion regions obtained using the following information: the CMS constraints on charged long-lived particles discussed at the end of Sec. \ref{lmix};  the XENON100 results \cite{Aprile:2011hi}  after having taken into consideration the threshold effects \cite{DelNobile:2012tx}; and the Higgs to two gamma constraints discussed in the subsection $h\rightarrow \gamma \gamma$ of Sec. \ref{direct}, where we have taken two different values of $\lambda_{HS}$.  

\vskip .5cm 
The plot shows that the extension of the SM presented here can be tested at the LHC while providing  a dark matter candidate interacting with ordinary matter via magnetic operators which can be simultaneously investigated or observed in dark 
matter experiments.  

\acknowledgments
 We gladly thank Eugenio Del Nobile for useful discussions and careful reading of the manuscript. 
 
 \newpage
\appendix

\section{Electromagnetic form factors and the magnetic moment }
\label{ap:1}

\begin{figure}[h!]
  \centering
  \includegraphics[width=0.8\textwidth]{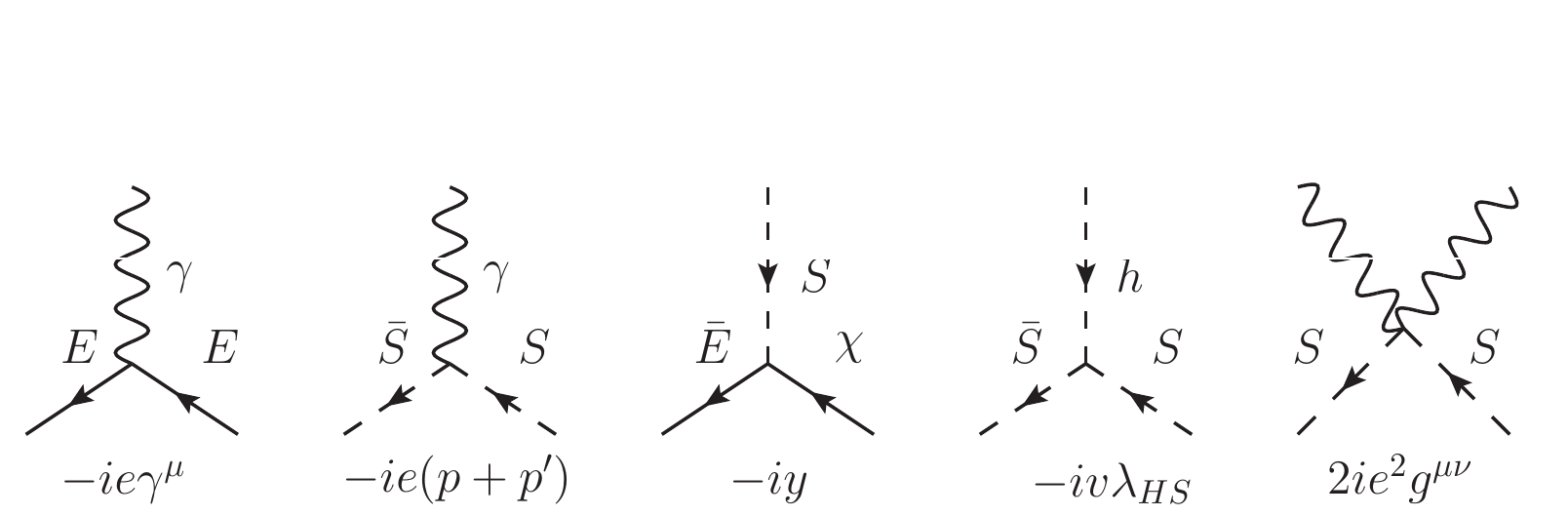}
  \caption{Feynman rules involving non-SM particles.}
  \label{fig:frules}
\end{figure}

To deduce the relevant processes as well as the magnetic moments for $\chi$, we review in Fig. \ref{fig:frules}
the Feynman rules to add to the SM which stem from our extension in \eqref{Sexy}.

The electromagnetic matrix element describing the interaction between a fermion and the photon can be written as
\begin{equation}
  \label{eq:EMelement}
  iT^\mu = -ie\bar{u}(p_1)\left[ \gamma^\mu F_1(q^2) + \frac{i\sigma^{\mu\nu}q_\nu}{2m_\chi}F_2(q^2) + \frac{i\sigma^{\mu\nu}\gamma^5q_\nu}{2m_\chi}F_3(q^2) \right]u(p_2).
\end{equation}
The second form factor corresponds to an anomalous magnetic moment of a fermion when evaluated in the static limit. We are interested in a neutral particle $\chi$, and, thus, the magnetic moment is determined by the second form factor
\begin{equation}
  \label{eq:mmdm}
  \vec{\mu}_{\chi} = g_\chi\left( \frac{e}{2m_\chi} \right)\vec{S} = \frac{\lambda_\chi}{2} \vec{S} ,
\end{equation}
where $g_\chi=2F_2(0)$. At zero momentum $F_1(0)=0$, and, therefore, there is no electric interaction between $\chi$ and the photon in this limit. 

The last term would give rise to an electric dipole moment for the $\chi$. Since the electric dipole moment is a $CP$-violating quantity and our theory does not possess new complex phases, the last form factor vanishes. The $CP$-violating phases in the Cabibbo-Kobayashi-Maskawa matrix will  provide an electric dipole moment for $\chi$ which is too small for the physical processes we discuss in this paper. The direct mixing with the SM leptons can induce also contributions to the electric dipole moments of $\chi$ which we will neglect here. 

\begin{figure}[htp]
  \centering
  \includegraphics[width=0.8\textwidth]{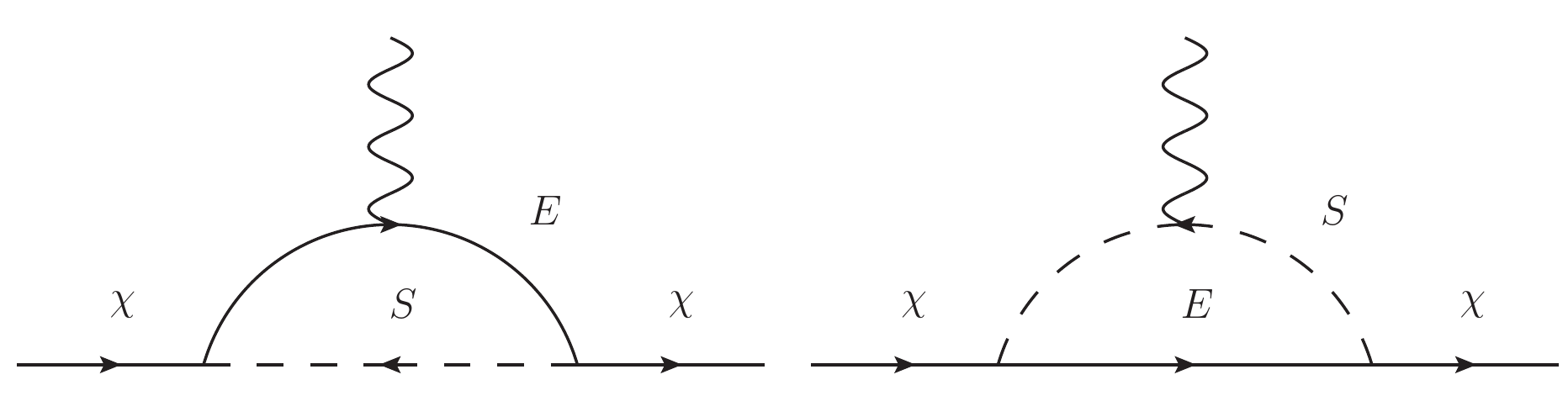}
  \caption{One-loop diagrams inducing the magnetic moment for $\chi$.}
  \label{fig:mmoneloop}
\end{figure}

To the one-loop level, two diagrams need to be computed to determine $F_1$ and $F_2$. They are drawn in Fig. \ref{fig:mmoneloop}.
The analytical expressions for the form factors $F_{1}(q^2)$ and $F_{2}(q^2)$ are  

\subsubsection*{First Diagram}
\begin{equation}
\begin{split}
&F'_{1,1}(q^2) =\int dx_1 dx_2 \frac{y^2}{(4\pi)^2}\\
 &\left[ \frac{2}{\epsilon^{\overline{\text{ms}}}}-1-\log\left(\frac{\Delta_1}{\mu^2} \right)+\frac{m_\chi^2(1-x_1-x_2)^2+2(1-x_1-x_2)m_\chi m_E+m_E^2+q^2x_1x_2}{\Delta_1} \right] \ ,
\end{split}
\end{equation}

\begin{equation}
F_{2,1}(q^2) =  \int dx_1 dx_2 y^2 \frac{2 m_\chi }{(4\pi)^2} \frac{(x_1+x_2)(m_\chi(1-x_1-x_2)+m_E)}{\Delta_1}  \ ,
\end{equation}

\begin{equation}
\Delta_1 = m_\chi^2(x_1+x_2-1)(x_1+x_2)-m_S^2(x_1+x_2-1)+m_E^2(x_1+x_2)-q^2x_1x_2 \ .
\end{equation}

\subsubsection*{Second Diagram}

\begin{equation}
\begin{split}
&F'_{1,2}(q^2) =-\int dx_1 dx_2 \frac{y^2}{(4\pi)^2}\\
& \left[ \frac{2}{\epsilon^{\overline{\text{ms}}}}-\log\left(\frac{\Delta_2}{\mu^2} \right)+  \frac{((x_2+x_1)m_\chi+m_E)(1-x_1-x_2))}{\Delta_2}2m_\chi\right] \ ,
\end{split}
\end{equation}

\begin{equation}
F_{2,2}(q^2) =  \int dx_1 dx_2 y^2 \frac{2 m_\chi }{(4\pi)^2} \frac{((x_2+x_1)m_\chi+m_E)(1-x_1-x_2)}{\Delta_2} \ ,
\end{equation}

\begin{equation}
\Delta_2 = (x_1+x_2)(m_\chi^2(x_1+x_2-1)+m_S^2)-m_E^2(x_2+x_1-1)-q^2x_1x_2 \ .
\end{equation}

We get the actual form factors by summing the contribution from both of the diagrams
\begin{align}
F_2(q^2)& = F_{2,1}(q^2)+F_{2,2}(q^2) \ ,\\
F_1(q^2)& =  F_{1,1}(q^2)+F_{1,2}(q^2)= F'_{1,1}(q^2)- F'_{1,1}(0)+F'_{1,2}(q^2)-F'_{1,2}(0) \\
	&=  F'_{1,1}(q^2)+\delta_1+F'_{1,2}(q^2)+\delta_2 \ . \nonumber
\end{align}
Here we have introduced one-loop counter terms, $\delta_1$ and $\delta_2$, which are fixed by the condition $F_{1,1}(0) = F_{1,2}(0) =0$.  Although this is redundant since the sum of the two contributions is finite, it will make the comparison easier when using the Cutkosky rules. Assuming $m_E=m_S=M$ and $q^2 << M^2$, we can expand the first form factor
\begin{equation}
F_1(q^2) = \frac{y^2q^2}{192 M^2 \pi^2} + \mathcal{O}(M^{-4}),
\end{equation}¥
and verify that it vanishes when the momentum goes to zero.

In order to reliably calculate the imaginary parts in the timelike region, $q^2>0$, we employ the Cutkosky rules. This allows us also to cross-check our calculations above because the real part is completely determined by the imaginary part through the un-subtracted dispersion relation ($s\equiv q^2$)
\begin{equation}
\operatorname{Re} F(s)=\frac{\mathcal{P}}{\pi}\int_{4 m_{E (S)}}^\infty \frac{\operatorname{Im} F(s')}{s'-s},
\end{equation}
or through the subtracted dispersion 
\begin{equation}
\operatorname{Re}F(s)=\frac{\mathcal{P}}{\pi}\int_{4 m_{E (S)}}^\infty \frac{s}{s'}\frac{\operatorname{Im} F(s')}{s'-s},
\end{equation}
when the form factor diverges in the static limit. Note that the sum of the contributions from both of the diagrams is finite at $q=0$ as expected for a dimension-5 operator. The analytical expressions for the imaginary parts are:

\begin{equation}
\begin{split}
 &\operatorname{Im} F_{1,1}= \\
& -\frac{4 \beta_E s}{256 \pi  s \large(s-2 m_{\chi }^2\large))^2} \large(2 m_{\chi }^2 \large(20 m_S^2+\large(5 \beta_E^2+4\large))
   s\large))+s \large(4 m_S^2+\large(\beta_E^2-3\large)) s\large))+16 m_e s m_{\chi }- \\
   & 4  m_{\chi }^4-32 m_e m_{\chi }^3\large))+\large(m_{\chi }^4 \large(-96 m_S^2+64 m_e^2+8
   \beta_E^2 s-36 s\large))-32 m_e m_{\chi }^3 \large(4 m_S^2+\large(\beta_E^2+2\large))
   s\large))+ \\
   & 2 m_{\chi }^2 \large(s \large(-32 m_e^2+5 \beta_E^4 s+3 s\large))+8 \large(5
   \beta_E^2+4\large)) s m_S^2+80 m_S^4\large))+16 m_e s m_{\chi } \large(4 m_S^2+\beta_E^2 s+s\large))+ \\
   &s \large(s \large(16 m_e^2+\large(\beta_E^2-1\large)){}^2 s\large))+8
   \large(\beta_E^2-1\large)) s m_S^2+16 m_S^4\large))+40 m_{\chi }^6+\\
   &64 m_e m_{\chi
   }^5\large)) \large(\log \large(-4 m_S^2+2 m_{\chi }^2-\large(\beta_E-1\large))^2
   s\large))-\log \large(-4 m_S^2+2 m_{\chi }^2-\large(\beta_E+1\large))^2
   s\large))\large)) \ ,
  \end{split}
\end{equation}

\begin{equation}
\begin{split}
 &\operatorname{Im} F_{1,2}=\\
  &  -\frac{m_{\chi } }{64 \pi  s \large(s-2 m_{\chi }^2\large)^2}\large(\large(m_{\chi } \large(-8 m_S^2 \large(3 \beta_E^2 s+s\large)-48
   m_S^4+\large(-3 \beta_E^4+2 \beta_E^2+1\large) s^2\large)+4 m_{\chi }^3 \large(4
   m_S^2- \\
   &\large(\beta_E^2+1\large) s\large)+8 m_e m_{\chi }^2 \large(4 m_S^2+\large(\beta_E^2-2\large) s\large)+4 m_e s \large(-4 m_S^2-\beta_E^2 s+s\large)+4 m_{\chi }^5+\\
   &16  m_e m_{\chi }^4\large) \large(\log \large(-4 m_S^2+2 m_{\chi }^2-\large(\beta_E-1\large)^2 s\large)-\log \large(-4 m_S^2+2 m_{\chi }^2-\large(\beta_E+1\large)^2 s\large)\large)-\\
   &4 \beta_E s \large(m_{\chi } \large(12 m_S^2+\large(3
   \beta_E^2-1\large) s\large)+4 m_e s+2 m_{\chi }^3-8 m_e m_{\chi
   }^2\large)\large) \ ,
   \end{split}
\end{equation}

\begin{equation}
\begin{split}
 &\operatorname{Im} F_{2,1}=\\
 &  \frac{1}{256 \pi  s
   \large(s-2 m_{\chi }^2\large)^2}\large(4 m_{\chi }^4 \large(8 m_e^2-\large(2 \beta_S^2+7\large) s\large)-32 m_e
   m_{\chi }^3 \large(4 m_e^2+\large(\beta_S^2+2\large) s\large)-10 m_{\chi }^2 \large(4
   m_e^2+\\ 
   &\beta_S^2 s+s\large) \large(4 m_e^2+\large(\beta_S^2-1\large) s\large)+16 m_e
   s m_{\chi } \large(4 m_e^2+\beta_S^2 s+s\large)-s \large(4 m_e^2+\large(\beta_S-1\large)^2 s\large) \large(4 m_e^2+ \large(\beta_S+ \\
   &1\large)^2 s\large)+24
   m_{\chi }^6+64 m_e m_{\chi }^5\large) \large(\log \large(2 m_{\chi }^2-4
   m_e^2-\large(\beta_S-1\large)^2 s\large)-\log \large(2 m_{\chi }^2-4
   m_e^2-\large(\beta_S+1\large)^2 s\large)\large)- \\
   &4 \beta_S s \large(2 m_{\chi }^2
   \large(20 m_e^2+\large(5 \beta_S^2-4\large) s\large)+s \large(4 m_e^2+\beta_S^2
   s+s\large)-16 m_e s m_{\chi }+12 m_{\chi }^4+32 m_e m_{\chi }^3\large) \ ,
\end{split}
\end{equation}

\begin{equation}
\begin{split}
& \operatorname{Im} F_{2,2}=\\
&  -\frac{m_{\chi } }{64 \pi  s \large(s-2 m_{\chi }^2\large)^2}\large(\large(m_{\chi } \large(-8 m_e^2 \large(3 \beta_S^2 s+s\large)-48
   m_e^4+\large(-3 \beta_S^4+2 \beta_S^2+1\large) s^2\large)+4 m_{\chi }^3 \large(4
   m_e^2- \\
   & \large(\beta_S^2+1\large) s\large)-8 m_e m_{\chi }^2 \large(4 m_e^2+\large(\beta_S^2+2\large) s\large)+4 m_e s \large(4 m_e^2+\beta_S^2 s+s\large)+4 m_{\chi }^5+\\
   &16   m_e m_{\chi }^4\large) \large(\log \large(2 m_{\chi }^2-4 m_e^2-\large(\beta_S-1\large)^2 s\large)-\log \large(2 m_{\chi }^2-4 m_e^2-\large(\beta_S+1\large)^2 s\large)\large)-\\
   &4 \beta_S s \large(m_{\chi } \large(12 m_e^2+\large(3
   \beta_S^2-1\large) s\large)-4 m_e s+2 m_{\chi }^3+8 m_e m_{\chi
   }^2\large)\large) \ .
\end{split}
\end{equation}

\section{Higgs Invisible Width}
\label{InvisibleH}
\begin{figure}[b!]
  \centering
  \includegraphics[width=0.5\textwidth]{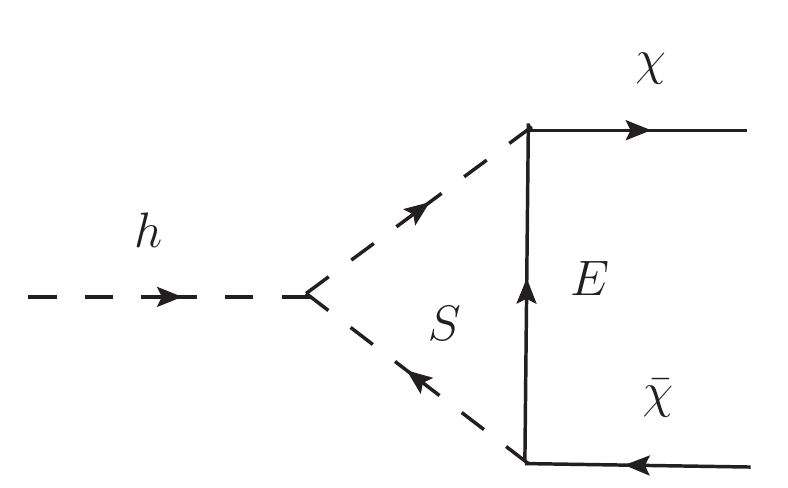}
  \caption{Invisible decay of the Higgs.}
  \label{fig:hdecay}
\end{figure}

The Higgs can decay to a pair of $\chi$s via loop-induced processes which at the one loop level is represented in Fig.~\ref{fig:hdecay}.
The formula for the decay width is
\begin{equation}
\Gamma = \frac{m_H}{32\pi}\left( 4 v^2\lambda_{HS}^2y^4 |A|^2  \right)\left( 1-\frac{4m_{\chi}^2}{m_H^2} \right)^{\frac{3}{2}}.
\end{equation}
The loop function $A$ can be expressed as
\begin{equation}
A=\frac{1}{(4\pi)^2}\int dx_1~dx_2~\frac{(x_1+x_2)m_{\chi}+m_E}{\Delta},
\end{equation}
where
\begin{equation}
\Delta = m_\chi^2 (x_1(x_1-1)+x_2(x_2-1))-2x_1x_2\left(\frac{m_H^2}{2} - m_\chi^2 \right)+(x_2+x_1)m_S^2+(1-x_1-x_2)m_E^2.
\end{equation}

\section{Monojet Cross Section}
\label{monoappendix}

Adopting a notation used in Ref. \cite{Fortin:2011hv}, the differential monojet cross section takes the form
\begin{equation}
\begin{split}
\frac{d^2\sigma^{q-\text{jet}}}{d x_{E_T}d\eta} = &\frac{\alpha^2\alpha_S}{36}\frac{se^\eta\Delta_1\Delta_2\Delta_3}{x_{m_\chi}^2}\left\{ \left| \frac{g_A}{c_w^2s(1-2x_{E_T}\cosh\eta)-M_Z^2+i\Gamma_Z M_Z} \right|^2 \right.+\\
&\left. \left| \frac{Q_q}{s(1-2x_{E_T}\cosh\eta)}+\frac{g_V}{c_w^2s(1-2x_{E_T}\cosh\eta)-M_Z^2+i\Gamma_Z M_Z} \right|^2\right\},
\end{split}
\end{equation}
\begin{equation}
\begin{split}
\frac{d^2\sigma^{g-\text{jet}}}{d x_{E_T}d\eta} = &\frac{4\alpha^2\alpha_S}{27}\frac{s\Delta_1\Delta_2\Delta_4}{x_{E_T}x_{m_\chi}^2}\left\{ \left| \frac{g_A}{c_w^2s(1-2x_{E_T}\cosh\eta)-M_Z^2+i\Gamma_Z M_Z} \right|^2 \right.+\\
&\left. \left| \frac{Q_q}{s(1-2x_{E_T}\cosh\eta)}+\frac{g_V}{c_w^2s(1-2x_{E_T}\cosh\eta)-M_Z^2+i\Gamma_Z M_Z} \right|^2\right\},
\end{split}
\end{equation}
where $Q_q$ is the charge of a quark, $\eta$ is the rapidity of a quark or a gluon, $g_V=\frac{1}{2}T_3-s_w^2Q_q$ and $g_A=-\frac{1}{2}T_3$. The dimensionless variables are defined as $\sqrt{s}x_{E_T}=E_T$  and $\sqrt{s}x_{m_\chi}=m_\chi$, $E_T$ denoting the transverse energy of a quark or a gluon. Finally, the expressions for $\Delta_i$ are
\begin{equation}
\begin{split}
\Delta_1 =& \left( 4x_{E_T}\cosh\eta\left( -4x_{m_\chi}^2\left( F_1^2+3F_1F_2+F_2^2 \right) +F_2^2 x_{E_T}\cosh\eta-F_2^2  \right) +\right. \\
&\left. 8x_{m_\chi}^2\left( F_1^2+3F_1F_2+F_2^2 \right) + 16F_1^2x_{m_\chi}^4 +F_2^2\right) ~, \\
\Delta_2 = & \sqrt{1-\frac{4m_{x_\chi}^2}{1-2x_{E_T}\cosh\eta}}~,\\
\Delta_3 = & x_{E_T}^2e^{-2\eta}+2x_{E_T}e^\eta(2x_{E_T}\cosh\eta-1)+1~,\\
\Delta_4 = & x_{E_T}^2\cosh(2\eta)-2x_{E_T}\cosh\eta+1 ~.
\end{split}
\end{equation}
From the gluon jet formula one can read off the monophoton differential cross section by multiplying the formula by the factor $\frac{9}{4}\frac{\alpha}{\alpha_S}$ and replacing  $Q_q$ with the electron charge.

\section{Higgs Decay to Two Photons}
\label{hgamma}

The Higgs decay width to two photons is modified by the presence of the new scalar state $S$ which interacts with both the Higgs and the photon. At one-loop level the decay width is
\begin{equation}
  \label{eq:hwidth}
    \Gamma=\frac{\alpha_{EW}^2G_Fm_h^3}{128\sqrt{2}\pi^3}\left|\sum_iN_{c,i}Q^2_iF_{f,i}+F_W+\left(\frac{2}{m_Sg}v\lambda_{HS}\right)F_S\right|^2,
\end{equation}
where $N_c$ is the number of colors, $Q$ is the charge of a particle, v is the vacuum expectation value and $g$ the weak coupling. Using the notation $\tau=\frac{4m^2}{m_h^2}$, the loop functions $F_i$ for fermions, bosons and scalars are given as
\begin{equation}
  \label{eq:loopF}
  \begin{split}
    F_W&=2+3\tau+3\tau(2-\tau)f(\tau),\\
    F_f&=-2\tau(1+(1-\tau)f(\tau)),\\
    F_S&=-\tau(1-\tau f(\tau)),
  \end{split}
\end{equation}
where
\begin{equation}
  \label{eq:ftau}
  f(\tau)=
  \begin{cases}
    \left(\arcsin{\sqrt{1/\tau}}\right)^2 & \text{if}\quad \tau \geq 1 \\
    -\frac{1}{4}\left[\log\left(\frac{1+\sqrt{1-\tau}}{1-\sqrt{1-\tau}}\right)-i\pi\right]^2 & \text{if} \quad\tau < 1.
  \end{cases}
\end{equation}

\end{document}